\documentclass[12pt,preprint]{aastex}

\usepackage{amsmath}

\slugcomment{ApJ In Press}

\shorttitle{SFH Systematic Uncertainties}
\shortauthors{Dolphin}

\begin{document}


\title{On the Estimation of Systematic Uncertainties of Star Formation Histories}


\author{Andrew E. Dolphin}
\affil{Raytheon Company, Tucson, AZ, 85734}
\email{adolphin@raytheon.com}

\begin{abstract}
In most star formation history (SFH) measurements, the reported uncertainties are those due to effects whose sizes can be readily measured: Poisson noise, adopted distance and extinction, and binning choices in the solution itself.  However, the largest source of error, systematics in the adopted isochrones, is usually ignored and very rarely explicitly incorporated into the uncertainties.  I propose a process by which estimates of the uncertainties due to evolutionary models can be incorporated into the SFH uncertainties.  This process relies on application of shifts in temperature and luminosity, the sizes of which must be calibrated for the data being analyzed.  While there are inherent limitations, the ability to estimate the effect of systematic errors and include them in the overall uncertainty is significant.  Effects of this are most notable in the case of shallow photometry, with which SFH measurements rely on evolved stars.
\end{abstract}





\section{Introduction \label{sec_intro}}

Resolved stellar populations provide a wealth of information from which a galaxy's star formation history (SFH) can be inferred.  The various features visible in a color-magnitude diagram (CMD) provide indicators of eras of star formation.  For example, the presence of upper main sequence stars indicates recent star formation, while horizontal branch stars indicate much older periods of star formation.  Since most epochs of star formation can be tied to specific populations, consideration of all age-sensitive populations permits a qualitative measurement of a galaxy's SFH, as summarized by \citet{hod89}.

More recently, quantitative approaches to this problem have been developed, in which the exact distribution of stars in a CMD is analyzed in order to more precisely determine the SFH \citep[e.g.,][]{gal96,tol96,dol97,her99,hol99,har01}.  These techniques rely on the use of synthetic CMDs that are generated by making use of theoretical isochrones, a stellar initial mass function, a model of observational effects, and often other factors such as reddening, foreground stars, and unresolved binaries.  These synthetic CMDs can be generated for an unlimited number of possible SFHs, and the history producing the synthetic CMD most resembling the observed data is reported as the measured SFH.

To make this comparison, the CMD is generally divided into a number of bins.  Within each bin, the number of observed stars is compared with the synthetic model using a goodness of fit statistic such as $\chi^2$, which can be written in the case of Poisson statistics as
\begin{equation}\label{eq_chi2}
\chi^2 = \sum \frac{(n_i - m_i)^2}{m_i},
\end{equation}
where $m_i$ is the number of model points in the bin and $n_i$ is the number of observed points.  A comparable statistic based on a Poisson noise model \citep{dol02} is given by
\begin{equation}\label{eq_logP}
-2 \ln P = 2 \sum m_i - n_i + n_i \ln \frac{n_i}{m_i}.
\end{equation}

With either statistic, the resulting figure of merit from the CMD comparison is related to the probability that the observed data were drawn from the synthetic model.  Assuming a uniform prior on all potential SFHs ($P(SFH)=1$), Bayes' theorem can be invoked, and this value can be interpreted as the probability that the SFH used to generate that synthetic model is correct:
\begin{equation}\label{eq_bayes}
P(SFH|CMD) = \frac{P(CMD|SFH)\ P(SFH)}{P(CMD)}.
\end{equation}

Uncertainties of the measured SFH are generally calculated with some form of Monte Carlo analysis.  This involves the generation of a large number of simulated observed CMDs, either from random realizations of the best-fitting model or from random resamplings of the original photometry.  The SFH is measured for each of these simulated CMDs, and the variation in these SFHs is interpreted as the uncertainty in the SFH of the original data.  Effects of uncertainties from distance, extinction, or the solution method itself can be incorporated as well.

A factor excluded from SFH analysis as described above is the effect of systematic uncertainties in the adopted isochrone set.  This raises several concerns about the method.  First, the possibity that observed data cannot be \textit{exactly} modeled by any combination of isochrones could potentially invalidate the use of a probabilistic technique for CMD comparison.  Second, if the probabilistic technique is capable of finding a well-constrained best match, the errors induced in that fit need to be understood and quantified.  The present study attempts to address these issues.

\section{Effects of Systematics on the Solution Space \label{sec_solve}}

To examine the effect of systematic errors on the ability to arrive at a solution, I have chosen the stressing case of extremely deep photometry (below the main sequence turnoff) of a simple stellar population.  (The term ``simple'' is used instead of ``single'' because the adaptations of isochrones used in this analysis contain spreads of 0.05 dex in age and 0.1 dex in metallicity.)  The depth of photometry requires that all parts of the CMD are adequately fit in order to achieve a good solution, while the simple stellar population assumption requires that all observed stars are modeled using a single isochrone (rather than allowing multiple isochrones to cover the entire observed data).  The latter requirement is particularly stressing on a probabilistic approach, since either formulation of the likelihood (Equations \ref{eq_chi2} and \ref{eq_logP}) becomes infinite if a CMD bin contains observed data but no model data.

The analysis was done by creating a synthetic population of stars with ages from 10.0 to 11.2 Gyr (log(age) of 10.00 to 10.05), metallicities from $-1.9$ to $-2.0$ in [Fe/H], and the Padua isochrones \citep{mar08} with updated AGB models \citep{gir10}.  Distance modulus and extinction were set to zero, and the $V$, $I$ filter combination was used.  The SFH was then measured using the MATCH package \citep{dol02}.  Three different isochrone sets were used for the solutions: the identical Padua models used to create the data, solar-scaled BaSTI models \citep{pie04} with $\eta=0.4$ and overshoot, and solar-scaled Dartmouth models \citep{dot08}.  Some amount of variability in RGB mass loss has been modeled in the BaSTI and Dartmouth synthetic CMDs, but not in those computed from the Padua models.  The solution was a four parameter fit for age, metallicity, distance, and extinction.  The age and metallicity parameters are the ones of primary scientific interest; the variations and distance and extinction allow for slight differences in the photometric calibrations adopted when computing the isochrones.

\begin{figure}
\epsscale{1.0}
\plotone{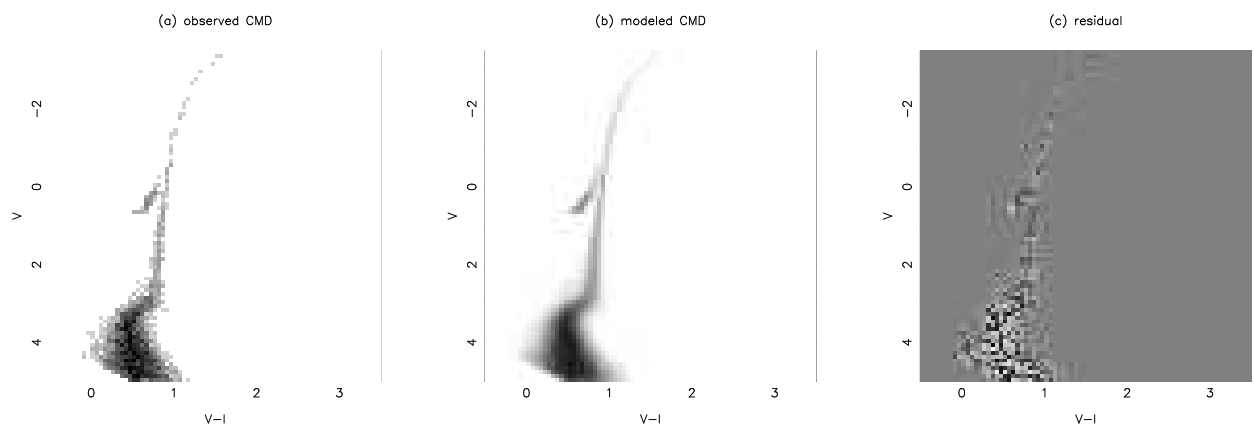}
\caption{Best-fit model CMD for a simple stellar population.  The left panel shows the simulated observed data, generated using the Padua isochrones \citep{mar08,gir10}.  The middle panel shows the best-fit model, also using the Padua isochrones.  The right panel shows the residual from subtracting the best-fit model from the simulated observations. \label{fig_paduaSSP}}
\end{figure}

\placefigure{fig_paduaSSP}
The best solution obtained using the Padua isochrones is shown in Figure \ref{fig_paduaSSP}.  There are no statistically significant differences between this solution and the synthetic population, and the best-fitting synthetic CMD was from the same population used to generate the data.  This is naturally the best-case scenario.

\begin{figure}
\epsscale{1.0}
\plotone{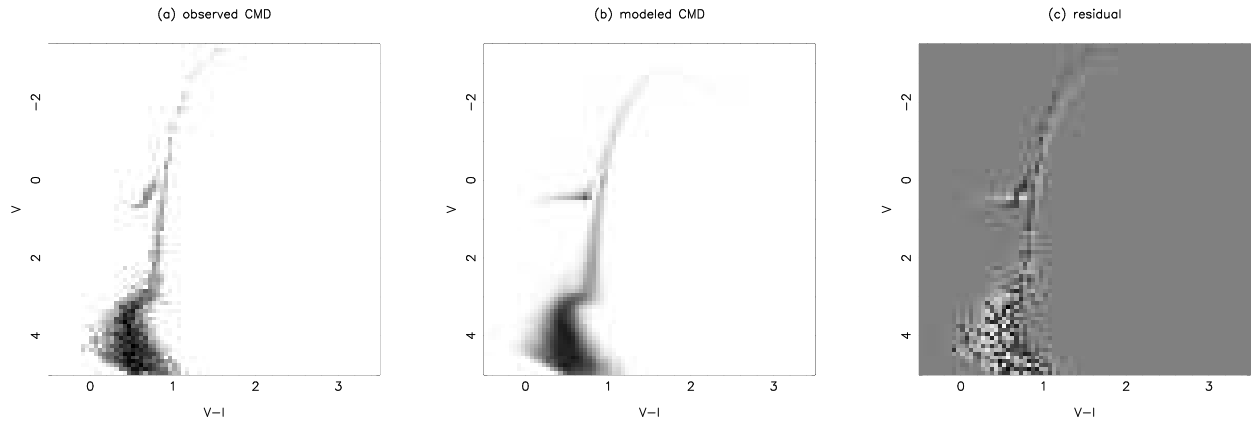}
\caption{Same as Figure \ref{fig_paduaSSP}, but using the BaSTI isochrones \citep{pie04} to fit the data. \label{fig_bastiSSP}}
\end{figure}

\placefigure{fig_bastiSSP}
The best solution obtained from the BaSTI isochrones is shown in Figure \ref{fig_bastiSSP}.  As expected, the fit is not nearly as clean as the previous solution.  Quantitatively, the synthetic photometry is $10^{150}$ times more likely to have been produced by the Padua models than by the BaSTI models.  Because no single BaSTI isochrone exactly matches the Padua isochrone used to generate the data, the algorithm must prioritize the CMD regions.  The choice is made to model the most strongly-populated parts of the CMD (the main sequence and subgiants) as well as possible, while allowing the upper RGB to deviate very strongly.  The horizontal branch is also fit as well as possible given the differences in the models.  The best fit is obtained with a slightly younger age ($7.9-8.9$ Gyr), higher metallicity ($-1.3$ to $-1.4$), and slight errors in both distance and extinction.

\begin{figure}
\epsscale{1.0}
\plotone{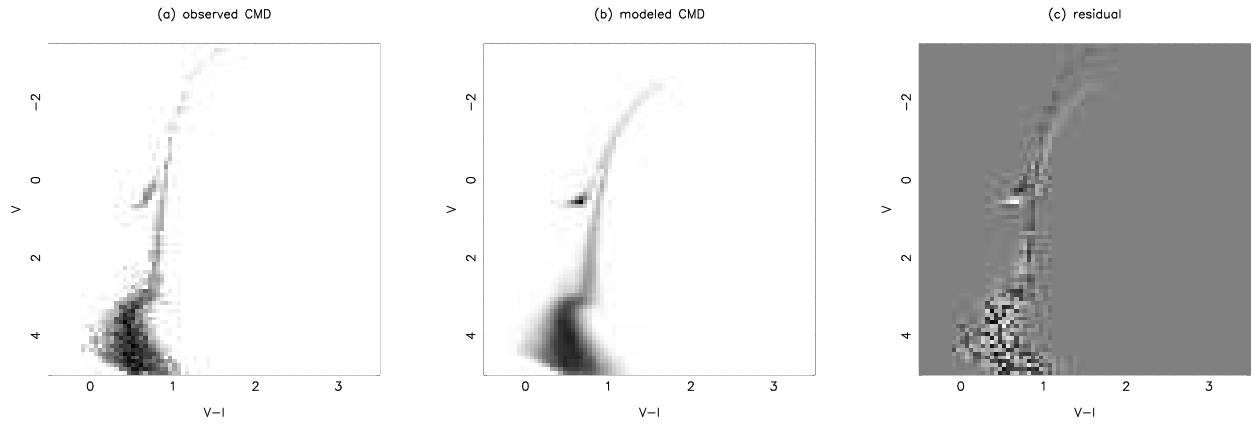}
\caption{Same as Figure \ref{fig_paduaSSP}, but using the Dartmouth isochrones \citep{dot08} to fit the data. \label{fig_dartmouthSSP}}
\end{figure}

\placefigure{fig_dartmouthSSP}
Finally, the best solution obtained from the Dartmouth isochrones is shown in Figure \ref{fig_dartmouthSSP}.  As with the BaSTI solution, there are significant errors in the fit, as again no single isochrone exactly matched the Padua models used to generate the simulated observations.  In this case, the synthetic photometry is $10^{106}$ times more likely to have been drawn from the Padua models than from the Dartmouth models -- an improved fit relative to the BaSTI models but nevertheless not a good match.  And, as with the BaSTI solution, the model producing the best fit did not match the input population; the stars were younger ($8.9-10$ Gyr) and significantly higher metallicity ($-1.0$ to $-1.1$).  The distance was nearly identical, but the extinction was 0.16 magnitudes less in $A_V$.

\placetable{tab_solve}
In addition to using the best solution to estimate the values of the four parameters, the dependencies of the goodness-of-fit on the four solution parameters can be used to estimate how well those parameters are constrained.  As the fit parameter from Equation \ref{eq_logP} is derived from the Poisson probability distribution, it can be converted directly into a four-dimensional probability density map.  Each parameter's probability density function can be obtained by marginalizing the other three.  Applying this procedure, the best fits and uncertainties from the three solutions above are listed in Table \ref{tab_solve}.

While there are clear systematic errors in the results shown in this table, it is encouraging that, even under stressing conditions, a probabilistic routine is able to measure a most probable set of parameters.  More significantly, the minimum surrounding the best solution is similarly well-defined in the presence of systematic errors as it is without them.  As seen in Table \ref{tab_solve}, the uncertainties in age and metallicity measured when using the BaSTI models are about twice those measured using the Padua models; the uncertainties in the Dartmouth-based solution are comparable to the Padua-based solution.

It should be noted that the behavior of a CMD-fitting routine in the face of such low probabilities likely depends strongly on the robustness of the algorithm.  However, at least in the case of the SFH measurement package being used in this analysis, the ability to find a solution is not significantly impeded by the presence of significant systematic errors.

\section{Simple Stellar Populations \label{sec_simple}}

With confidence that precise (though inaccurate) fits can be obtained despite the presence of systematic errors, a logical step is to quantify the size of the errors in measured stellar populations induced by these errors in the models used to measure them.  To accomplish this, the experiment described in the previous section was expanded.  All three isochrone sets were used to generate the synthetic photometry, and five populations of different ages and metallicities were simulated.

\placetable{tab_simple}
Results from this experiment are shown in Table \ref{tab_simple}.  Looking at the summary lines, it is clear that the effects of systematic differences between the three isochrone sets is not constant between the populations.  For example, there were no errors in the youngest population's age measurement, but the metallicity was in error by $0.2$ dex in both cases.  On the other hand, the $\sim 2.5$ Gyr population had a smaller error in metallicity but a larger error in age.

Because of the non-uniformity of age and metallicity systematic uncertainties, it is tempting to draw the conclusion that uncertainties should be estimated by measuring the SFH with as many isochrone sets as is possible, and setting the uncertainty equal to the standard deviation of the solutions.  While this is possible in some cases, there are two significant limitations.  First, not all isochrone sets will cover the necessary ranges in age, metallicity, and evolutionary phases.  For example, of the models used in this analysis, only the Padua models provide isochrones younger than 25 Myr.  Thus, systematic errors could not be measured for systems containing ongoing star formation.

Even in the case of older populations for which all isochrone sets used here are available, the use of three SFH measurements to estimate the uncertainty in created by systematic errors is analogous to using two Monte Carlo runs to estimate random errors.

\section{Estimation of Systematic Uncertainties \label{sec_errors}}

In order to estimate systematic uncertainties, one must be able to generate nearly unlimited representations of reasonable systematic errors, each of which can be analyzed.  This section outlines one such method, although other viable alternatives certainly exist.  It should be emphasized that the primary motivation of this study is to understand and quantify effects of model differences on the measured SFH, not to examine or critique the underlying modeling choices and methods that created these differences.

The adopted approach is to model the isochrone differences as shifts in $M_{bol}$ and $\log T_{eff}$.  The rationale behind the choice of $\log T_{eff}$ rather than color is that a simple color shift could result in unphysical temperatures (e.g., $V-I=-1$).  It also produces a system that is more likely to be portable between different filter sets with minimal modifications.

It should be emphasized that this is not to be confused with shifts in distance modulus and extinction that are commonly applied during the determination of the best-fit star formation history (including the solutions shown in Table \ref{tab_solve}).  The distance modulus and extinction shifts are used to compensate for uncertainties such as those in the photometric zero points, photometric calibration of the isochrone sets, and foreground extinction.  The shifts described in this section are used to intentionally introduce systematic errors between the observed data (which is not shifted) and the synthetic models (which are).  It is important to note that, for solutions in which these shifts are applied, no solution in distance modulus or extinction is allowed, since this would mitigate the inserted error.  Specifically, any error induced in the bolometric magnitudes could exactly be offset by an identical shift in distance modulus, thereby resulting in exactly the same star formation history as would have been observed had no shifts been applied to either variable.

For this to work successfully, it is necessary that a set of shifts in the $M_{bol}$, $\log T_{eff}$ space can adequately reproduce the errors in age and metallicity for all five populations shown in Table \ref{tab_simple}.  In order to make this transformation, the change in recovered age and metallicity can be measured as a function of the shifts in luminosity and temperature.  For the youngest population in Table \ref{tab_simple}, this can be described with the following equation:
\begin{equation}\label{eq_matrix}
\begin{pmatrix}
\Delta \log(age) \\
\Delta [Fe/H]
\end{pmatrix} =
\begin{pmatrix}
-0.25 & -0.05 \\
13.5  &  1.2
\end{pmatrix}
\begin{pmatrix}
\Delta \log T_{eff} \\
\Delta M_{bol}
\end{pmatrix}.
\end{equation}

This indicates that the accurate recovery of the age of the population (even in the presence of systematic errors) is because it would require a very large error in temperature or luminosity to create a noticeable change in the measured age.  However, a relatively small difference in either temperature or luminosity could create a significant error in the recovered metallicity (a result of the relative insensitivity of upper main sequence color to metallicity).

One can invert the matrix in Equation \ref{eq_matrix} to solve for the temperature and luminosity shifts that would result in the observed errors in age and metallicity.  In this example, there was zero age error but $-0.2$ dex of metallicty error when generating data with the Padua models and solving with the BaSTI models.  Put into the above equation, this translates to temperature and luminosity shifts of
\begin{equation}\label{eq_matrix_inv}
\begin{pmatrix}
\Delta \log T_{eff} \\
\Delta M_{bol}
\end{pmatrix} =
\begin{pmatrix}
3.2 &  0.13 \\
-36 & -0.67
\end{pmatrix}
\begin{pmatrix}
0.0 \\
-0.25
\end{pmatrix} =
\begin{pmatrix}
-0.033 \\
0.17
\end{pmatrix}
\end{equation}

Indeed, when applying shifts of $-0.027$ in $\log T_{eff}$ and $0.133$ in $M_{bol}$ while solving the SFH using the Padua models on data generated with the Padua models, one induces the expected error in metallicity, with zero age error.  Expanding this result, many solutions could be made with shifts comparable in size to these.  In doing so, a fuller sampling of the effects of systemtics can be obtained.

Recalling that it is necessary to find shifts that adequately model all of the populations in Table \ref{tab_simple}, similar math is performed on the other four.  The most extreme case is the oldest population, in which the matrix is nearly singular and thus very large variations of $\sigma_{\log T_{eff}} = 0.142$ and $\sigma_{M_{bol}} = 1.14$ are obtained.  However, the uncertainties on those values are also very high.  When combining the results of all five populations, values of $\sigma_{\log T_{eff}} = 0.012$ and $\sigma_{M_{bol}} = 0.18$ are found.

\placetable{tab_multiple}
Table \ref{tab_multiple} shows the resulting age and metallicity errors obtained by applying shifts randomly selected from a Gaussian distribution of mean zero and standard deviation as specified above.  While these errors do not exactly match those measured from the simple populations, they are well within a factor of two in all cases, and within $30 \%$ for all but one.  To put into context, MATCH's maximum resolutions in age and metallicity are 0.05 and 0.1 dex, respectively.  More importantly, in cases in which only two isochrones were used, the uncertainty of the standard deviation is $\sim 75 \%$ of the measured standard deviation.  Even in the cases for which three isochrone sets were used, the uncertainty is about half the measured standard deviation.  Thus, in the presence of at minimum $50 \%$ uncertainties in the measured error distributions, the results from Table \ref{tab_multiple} show the temperature and luminosity shifts to adequately reproduce the observed systematic errors in all five populations.

\section{Multiple Stellar Populations \label{sec_multiple}}

While the above section demonstrated that consistent shifts in temperature and luminosity can induce appropriately-sized errors in recovered simple populations across a wide spectrum of age, it needs to be seen whether or not the process would work for more complex systems such as field populations of resolved galaxies.  In this section, this topic is addressed by analysis of the two extreme cases: a population whose entire SFH is comprised of a small number of short bursts, and a population with constant star formation rate for its entire history.

For the first case, a four-burst stellar population was created using the BaSTI isochrones.  The four bursts were defined as follows:
\begin{itemize}
\item $50-56$ Myr, mean [Fe/H] = $-0.45$
\item $500-562$ Myr, mean [Fe/H] = $-0.75$
\item $2.5-2.8$ Gyr, mean [Fe/H] = $-1.15$
\item $12.6-14.1$ Gyr, mean [Fe/H] = $-1.95$
\end{itemize}

The SFH of this population was then measured using the Padua isochrones, with no shifts in temperature or luminosity.  In addition, 50 solutions were made with randomly chosen shifts (using Gaussian distributions with $\sigma_{\log T_{eff}} = 0.012$ and $\sigma_{M_{bol}} = 0.18$).  The variation between these solutions was used measure the systematic uncertainties.  Finally, to understand what the measurement would have been in the absence of systematic errors, the history was measured using the BaSTI isochrones.

\begin{figure}
\epsscale{1.0}
\plotone{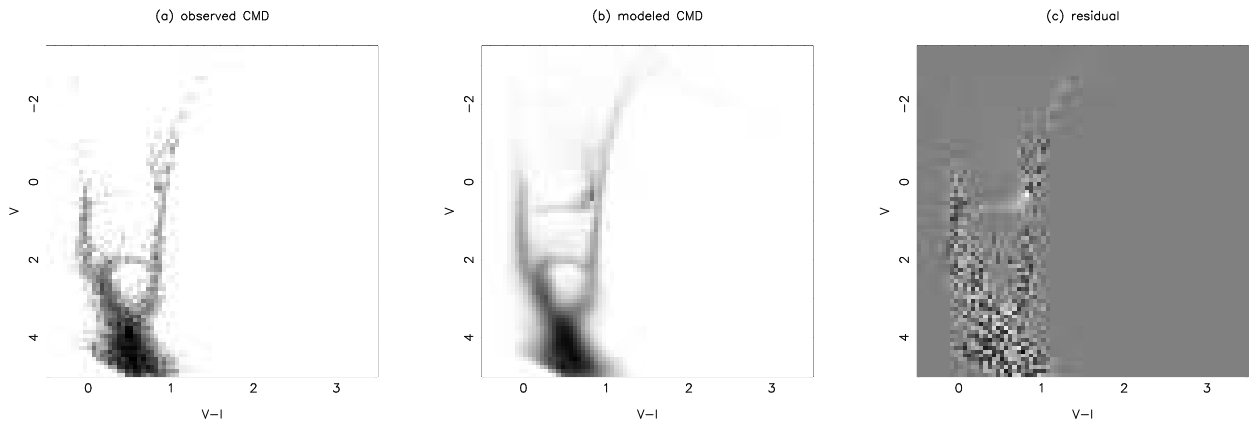}
\caption{Best-fit model CMD for a multi-burst stellar population.  The BaSTI models were used to generate the simulated observations, and the Padua models were used to fit the data.  Panels are the same as in figure \ref{fig_paduaSSP}. \label{fig_multi_burst_cmd}}
\end{figure}

\placefigure{fig_multi_burst_cmd}
The best fit using the Padua models is shown in Figure \ref{fig_multi_burst_cmd}, and is not nearly as bad as the fits shown in Section \ref{sec_solve}.  The reason for this is that MATCH is allowed to mix populations, allowing more portions of the CMD to be fit.  In this case, only the horizontal branch was fit poorly.

\begin{figure}
\epsscale{1.0}
\plotone{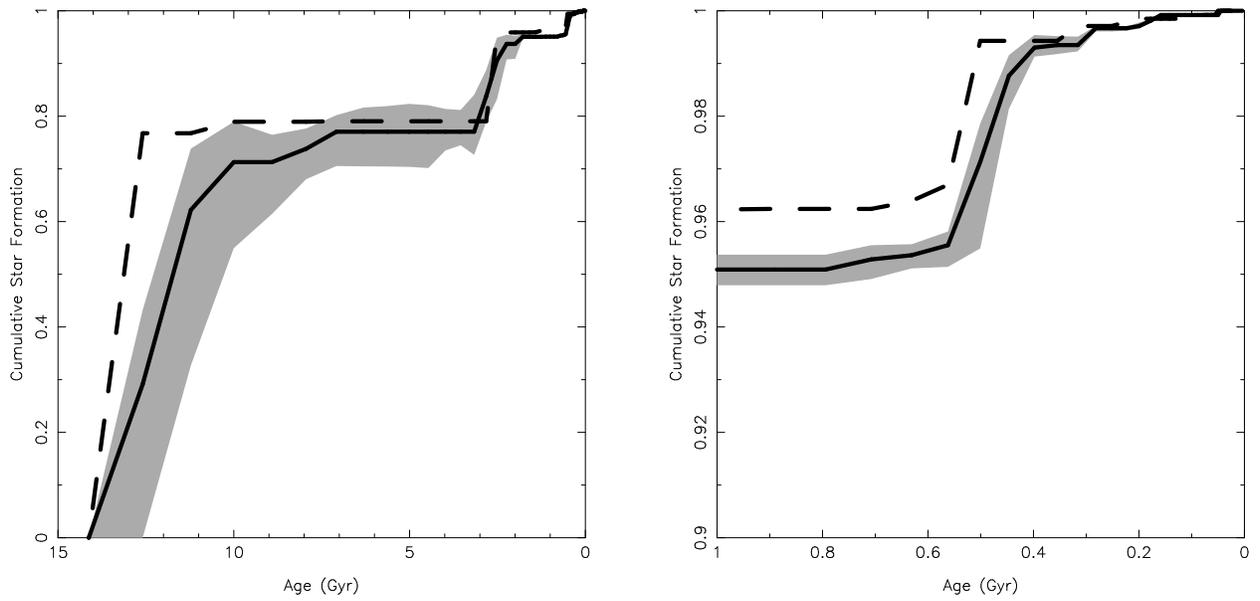}
\caption{Measured SFH of a stellar population with four bursts of star formation, in the presence of systematic errors.  The solid line represents the best fit to the SFH, and the shaded region shows the one-sigma uncertainties.  Finally, the history as measured with no systematic errors is shown by the dashed line.  The left panel shows the entire history of the galaxy; the right panel zooms into the past Gyr. \label{fig_multi_burst_sfh}}
\end{figure}

\placefigure{fig_multi_burst_sfh}
The measured SFHs are shown in Figure \ref{fig_multi_burst_sfh}.  The solid line and shaded region show the best fit using the Padua models and the uncertainty estimated from the fits with the Padua models and shifts in luminosity and temperature.  The dashed line shows the best fit obtained using the BaSTI models.  The error-free measurement generally sits near the upper error bars, though at some times (for example, from $0.5$ to $2.5$ Gyr ago) falls significantly outside the bounds.  In this case, the ratio of star formation in the middle two bursts was systematically mis-estimated in all of the runs with the Padua models.  Most likely this is a result of the two isochrone sets having different lifetimes of certain evolutionary phases, an error source not accounted for in this analysis.  On the other hand, the estimated uncertainties on the measured burst ages were correct.

The second test was carried out in the other extreme, a constant star formation rate from 14.1 Gyr ago to the present.  As before, the synthetic data were generated using the BaSTI models and solved using the Padua models, including 50 runs with temperature and luminosity shifts to estimate the systematic uncertainties.

\begin{figure}
\epsscale{1.0}
\plotone{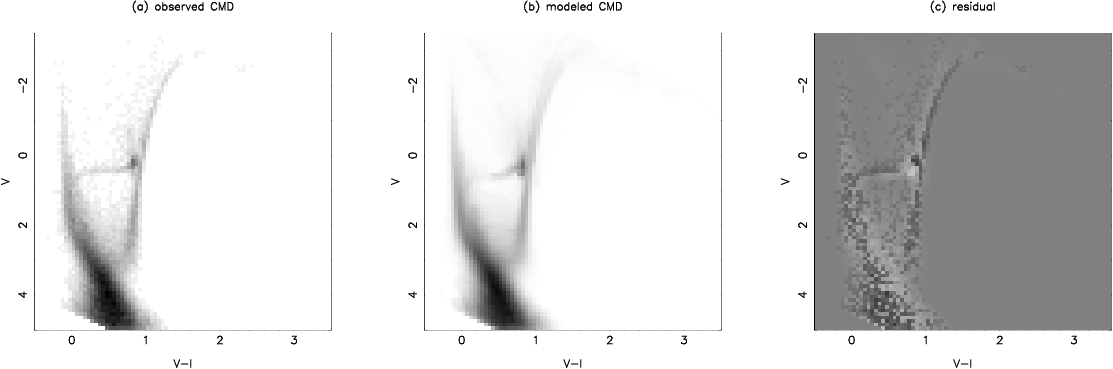}
\caption{Best-fit model CMD for a stellar population with continuous star formation.  The BaSTI models were used to generate the simulated observations, and the Padua models were used to fit the data.  Panels are the same as in Figure \ref{fig_paduaSSP}. \label{fig_multi_const_cmd}}
\end{figure}

\begin{figure}
\epsscale{1.0}
\plotone{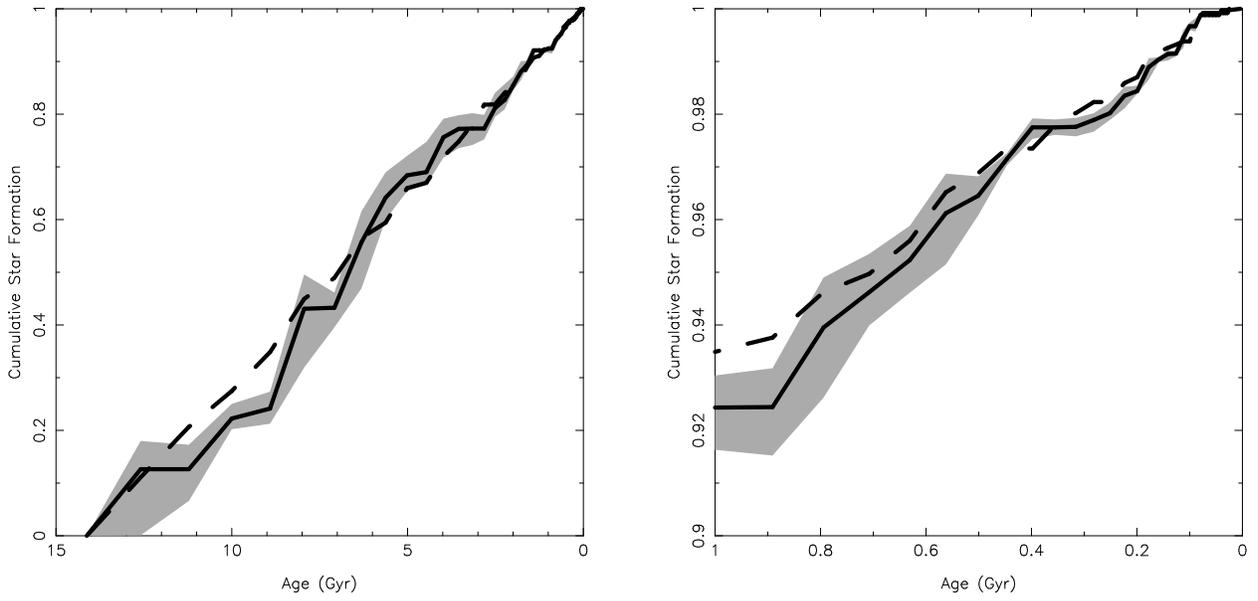}
\caption{Measured SFH of a stellar population with continuous star formation, in the presence of systematic errors. Panels and lines are identical to those in Figure \ref{fig_multi_burst_sfh}. \label{fig_multi_const_sfh}}
\end{figure}

\placefigure{fig_multi_const_cmd}
\placefigure{fig_multi_const_sfh}
Results of this test are shown in Figures \ref{fig_multi_const_cmd} and \ref{fig_multi_const_sfh}.  The quality of the fit was not nearly as good as in the multiple-burst test, as the presence of all ages required all CMD features to be fit. The estimation of the uncertainties appears to have been successful; the measurement with no systematic errors fell generally within the error bars and never outside them by much.  The largest deviation, seen at $8.9$ Gyr, is equivalent to a two-sigma error.

\section{Discussion \label{sec_discussion}}

Ideally, there would be a nearly limitless number of suitable isochrone sets covering the entire space of reasonable choices of the modeling parameters.  The reality is that one must make the best possible use of a handful of models, not all of which may cover the entire range of initial mass or evolutionary phases that are needed in an analysis.  Thus, the direct measurement of systematic errors by analysis of multiple isochrone sets is inaccurate at best, and impossible at worst.

The approach I have outlined for estimating the uncertainties due to systematic errors has relied on finding some simple proxy for the isochrone differences that can be varied in a large number of Monte Carlo runs in order to observe its effects.  This provides the advantage that it can be applied in populations for which only one isochrone set provides coverage (for the three isochrone sets analyzed here, this would be for ages younger than 25 Myr), and also that rather than relying on one or two independent pairs of isochrone sets to sample the distribution of errors produced in the SFH measurement, a nearly unlimited sample can be generated by random draws.

The choice made was to use shifts in luminosity and effective temperature, which can be applied relatively simply during the CMD fitting process.  This ignores systematic differences in measured histories that could result from different lengths of evolutionary phases.  In cases in which the SFH is dominated by a small number of short bursts, this can cause the uncertainties in the ratio of star formation in each burst to be underestimated.  For field populations with continuous star formation, this is less of a problem.

\placetable{tab_photometric_depth}
The measurement of appropriate shift sizes is thus key to the success.  Because the systematic error model is not physics-based, the shifts should be measured for every data set.  For example, photometric depth will affect which CMD features dominate the measurement of the SFH, causing the best overall shift size to change somewhat based on how sensitive those features are to modeling choices.  Effects of this on the shift sizes are shown in Table \ref{tab_photometric_depth}.  Data shown in this table are computed using the averages of five measurements of the shifts at each photometric depth.  While the naive expectation might have been a reduction in shifts as the photometry is deeper, this is not true in every case.  What is true is that the resulting errors in age and metallicity become progressively smaller as the photometry becomes deeper.

\placetable{tab_filters}
\placetable{tab_photometric_depth_bv}
\placetable{tab_photometric_depth_jk}
In addition to photometric depth, the filter choice and stellar population being examined can affect the ideal shifts to apply to a data set.  Table \ref{tab_filters} shows the shifts applied to deep photometry (beyond the ancient main sequence turnoff) for three different filter combinations.  While the $B,V$ filter combination behaves similarly to $V,I$, much larger luminosity variations are needed in $J,K$.  Population effects can also be present.  As seen in Section \ref{sec_simple}, the same shifts are not obtained for every age or metallicity.  Thus, the appropriate shifts must be measured for the specific conditions seen in the data being analyzed.  For the sake of completeness, shifts as a function of photometric depth for the $B,V$ filter combination is given in Table \ref{tab_photometric_depth_bv}, and in Table \ref{tab_photometric_depth_jk} for the $J,K$ combination.

\begin{figure}
\epsscale{1.0}
\plotone{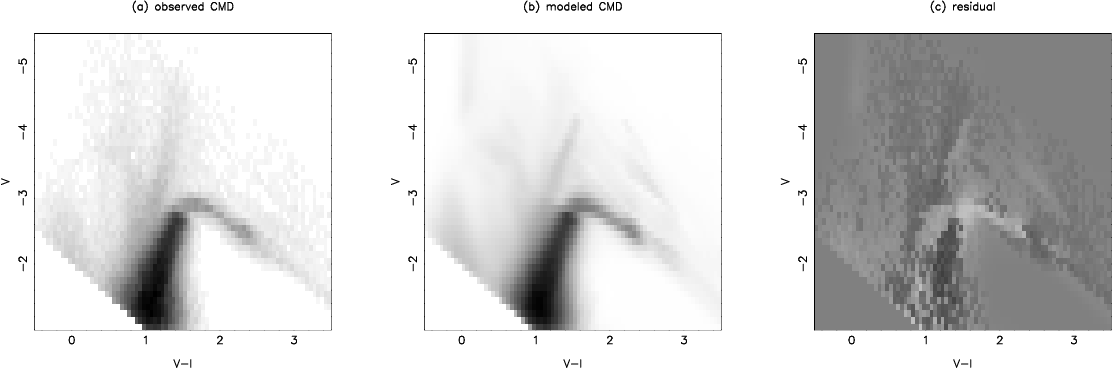}
\caption{Best-fit model CMD for a stellar population with continuous star formation and shallow photometry.  The BaSTI models were used to generate the simulated observations, and the Padua models were used to fit the data.  Panels are the same as in Figure \ref{fig_paduaSSP}. \label{fig_shallow_cmd}}
\end{figure}

\begin{figure}
\epsscale{1.0}
\plotone{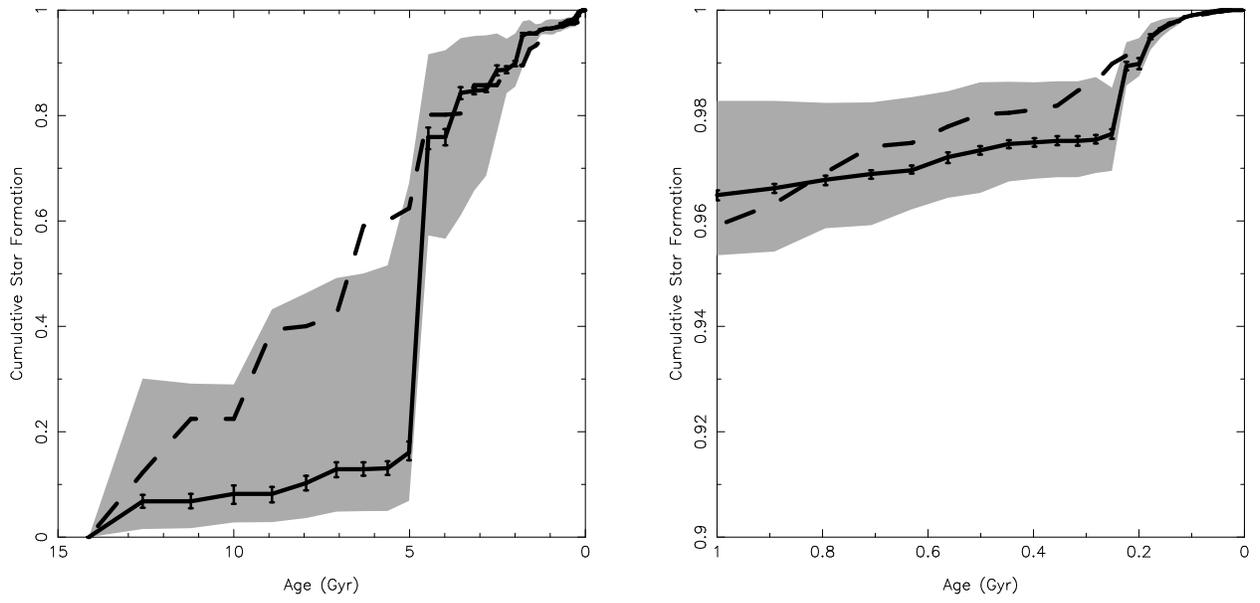}
\caption{Measured SFH of a stellar population with continuous star formation, in the presence of systematic errors. Panels and lines are identical to those in Figure \ref{fig_multi_burst_sfh}, except that uncertainties due to random errors (shown as the error bars) have been added. \label{fig_shallow_sfh}}
\end{figure}

\placefigure{fig_shallow_cmd}
\placefigure{fig_shallow_sfh}
The goal of this analysis is a reliable method for estimating systematic uncertainties that can be incorporated into SFH results.  In some cases, this significantly increases the uncertainties and resolves issues in which SFH algorithms report error bars that appear unrealistically small.  For example, the CMD shown in Figure \ref{fig_shallow_cmd} contains very little information for older populations: the bright red giants and asymptotic giant branch stars, for which not only are significant age/metallicity degeneracies present, but also the evolutionary modeling is less certain.  The measured SFHs from this field are shown in Figure \ref{fig_shallow_sfh}.  The systematic error estimate (grey region) shows the expected large uncertainty, while random errors only (error bars) are extremely small, and incorrectly show that approximately $70 \%$ of the star formation to have occured in a sharp burst 5 Gyr ago.  When incorporating systematic errors, the lack of age resolution is apparent and no statistically significant burst is measured.  The solution with no systematic errors is thus in error at more than a ten-sigma level if using random errors only, while it is within the estimated systematic uncertainties at most ages.

\section{Summary \label{sec_summary}}

Systematic uncertainties, in the form of uncertainties in the adopted isochrones, are a significant (and frequently dominant) source of uncertainty in the measurement of SFHs.  Despite this, SFH measurements are generally reported with minimal (if any) analysis indicating the potential effects of systematic errors on the results.


I have outlined a process for estimating the effects of systematic errors on the measured SFH.  This process is based on shifts in luminosity and temperature that can be applied.  The shifts can be varied within some probability distribution, allowing the estimation of a large number of SFHs with reasonably sized systematic errors.

Being an entirely empirical technique, there are limitations.  First, the sizes of the shifts to apply are dependent on the stellar population being observed, the filter set in use, and the depth of the photometry.  The implication is that the shifts will be data set dependent, and thus should be re-calculated for each data set.  Second, while the method was calibrated to measure errors in inferred age and metallicity to reasonable accuracy, effects due to lifetimes of certain evolved populations are not an explicit part of the calibration.  For field populations, this is not seen to cause significant issues.  However, uncertainties in the histories of populations formed by a small number of short bursts can be underestimated.

Limitations notwithstanding, the analysis presented here indicates a significant improvement in the estimation of uncertainties in SFHs.  The ability to obtain extremely small random errors in shallow, wide fields has long been known \citep[e.g.,][]{dol02}, simply due to the effects of sample size on random errors.  Inclusion of systematic errors in the uncertainty analysis prevents this, and results in error bars that more accurately represent the degree of confidence in the measured SFH.

\acknowledgments

Support for program number HST-GO-11986.07 was provided by NASA through a grant from the Space Telescope Science Institute, which is operated by the Association of Universities for Research in Astronomy, Incorporated, under NASA contract NAS5-26555.

\clearpage

\clearpage

\clearpage

\clearpage

\clearpage

\clearpage

\clearpage

\clearpage

\clearpage

\clearpage

\begin{deluxetable}{lrrrr}
\tablewidth{0pt}
\tablecaption{Measured parameters of a simple stellar population.  The data were simulated using the Padua models; three different model sets were used to measure the SFHs.\label{tab_solve}}
\tablehead{
\colhead{Data Set} &
\colhead{Age} &
\colhead{Metallicity} &
\colhead{Distance} &
\colhead{Extinction} \\
\colhead{} &
\colhead{(Gyr)} &
\colhead{([Fe/H])} &
\colhead{$m-M$} &
\colhead{($A_V$)}
}
\startdata
Input Population   & 10.59 & -1.95 & 0.000 & 0.000 \\
Padua Solution     & $10.73 \pm 0.07$ & $-1.948 \pm 0.015$ & $0.003 \pm 0.004$ & $-0.005 \pm 0.003$ \\
BaSTI Solution     &  $8.90 \pm 0.10$ & $-1.318 \pm 0.033$ & $0.082 \pm 0.003$ & $-0.090 \pm 0.002$ \\
Dartmouth Solution &  $9.46 \pm 0.09$ & $-1.080 \pm 0.013$ & $0.011 \pm 0.004$ & $-0.161 \pm 0.003$ \\
\enddata
\end{deluxetable}

\clearpage

\begin{deluxetable}{lrrlrr}
\tablewidth{0pt}
\tablecaption{Best-fitting models for various simple stellar populations.  The left columns indicate the isochrone set and stellar population used to generate the CMD.  The right columns indicate the isochrone set used to solve for the best fit, as well as the mean age and mean metallicity of that best fit.  Simple stellar populations were used in both CMD generation and SFH measurement.  Note that standard deviations were estimated using the mean absolute value of the differences, which is an unbiased estimator of standard deviation.  For the case of a Gaussian distribution, the standard deviation equals the mean error multiplied by $\sqrt{\pi/2}$.  Note also that Dartmouth isochrones were not used at ages younger than 1 Gyr, as the younger isochrones are not implemented in MATCH.\label{tab_simple}}
\tablehead{
\multicolumn{3}{c}{Input} &
\multicolumn{3}{c}{Solution} \\
\colhead{Isochrones} &
\colhead{Log Age} &
\colhead{[Fe/H]} &
\colhead{Isochrones} &
\colhead{Log Age} &
\colhead{[Fe/H]}
}
\startdata
Padua        & 7.825 & -0.75 & BaSTI     & 7.825 & -0.95 \\
BaSTI        & 7.825 & -0.75 & Padua     & 7.825 & -0.55 \\
Standard Dev &       &       &           & 0.00 &  0.25 \\
\tableline
Padua        & 8.425 & -0.75 & BaSTI     & 8.425 & -0.55 \\
BaSTI        & 8.425 & -0.75 & Padua     & 8.375 & -0.85 \\
Standard Dev &       &       &           & 0.03 &  0.19 \\
\tableline
Padua        & 9.425 & -0.85 & BaSTI     & 9.375 & -0.65 \\
Padua        & 9.425 & -0.85 & Dartmouth & 9.525 & -0.65 \\
BaSTI        & 9.425 & -0.85 & Padua     & 9.475 & -1.05 \\
BaSTI        & 9.425 & -0.85 & Dartmouth & 9.475 & -0.85 \\
Dartmouth    & 9.425 & -0.85 & Padua     & 9.425 & -1.15 \\
Dartmouth    & 9.425 & -0.85 & BaSTI     & 9.375 & -0.85 \\
Standard Dev &       &       &           & 0.06 &  0.15 \\
\tableline
Padua        & 9.725 & -1.05 & BaSTI     & 9.725 & -1.05 \\
Padua        & 9.725 & -1.05 & Dartmouth & 9.725 & -0.95 \\
BaSTI        & 9.725 & -1.05 & Padua     & 9.725 & -1.05 \\
BaSTI        & 9.725 & -1.05 & Dartmouth & 9.775 & -1.15 \\
Dartmouth    & 9.725 & -1.05 & Padua     & 9.675 & -0.85 \\
Dartmouth    & 9.725 & -1.05 & BaSTI     & 9.675 & -0.85 \\
Standard Dev &       &       &           & 0.03 &  0.13 \\
\tableline
Padua        & 10.025 & -1.35 & BaSTI     &  9.975 & -0.95 \\
Padua        & 10.025 & -1.35 & Dartmouth & 10.025 & -0.85 \\
BaSTI        & 10.025 & -1.35 & Padua     & 10.025 & -1.65 \\
BaSTI        & 10.025 & -1.35 & Dartmouth & 10.075 & -1.45 \\
Dartmouth    & 10.025 & -1.35 & Padua     & 10.075 & -1.65 \\
Dartmouth    & 10.025 & -1.35 & BaSTI     &  9.975 & -1.45 \\
Standard Dev &        &       &           &  0.04 &  0.36 \\
\enddata
\end{deluxetable}

\clearpage

\begin{deluxetable}{rrrrrr}
\tablewidth{0pt}
\tablecaption{Comparison between measured and estimated errors due to systematic errors.  The first two columns show the age and metallicity of the population used to generate the data.  The third and fourth columns show the measured standard deviation in age and metallicity errors, and are repeated from Table \ref{tab_simple}.  The final two columns show the errors in age and metallicity obtained by applying random shifts with standard deviations of $\sigma_{\log T_{eff}} = 0.012$ and $\sigma_{M_{bol}} = 0.18$.\label{tab_multiple}}
\tablehead{
\multicolumn{2}{c}{Input} &
\multicolumn{2}{c}{Measured} &
\multicolumn{2}{c}{Calculated} \\
\colhead{Log Age} &
\colhead{[Fe/H]} &
\colhead{$\sigma_{\log\ age}$} &
\colhead{$\sigma_{[Fe/H]}$} &
\colhead{$\sigma_{\log\ age}$} &
\colhead{$\sigma_{[Fe/H]}$}
}
\startdata
 7.825 & -0.75 & 0.00 &  0.25 & 0.01 & 0.27 \\
 8.425 & -0.75 & 0.03 &  0.19 & 0.00 & 0.25 \\
 9.425 & -0.85 & 0.06 &  0.15 & 0.08 & 0.18 \\
 9.725 & -1.05 & 0.03 &  0.13 & 0.04 & 0.18 \\
10.025 & -1.35 & 0.04 &  0.36 & 0.03 & 0.21 \\
\tableline
\enddata
\end{deluxetable}

\clearpage

\begin{deluxetable}{rrrr}
\tablewidth{0pt}
\tablecaption{Variation of temperature and luminosity shifts as a result of variations in photometric depth.  The left two columns show the 50\% completeness limit; the right two columns show the shifts needed to adequately model effects of systematic erors. \label{tab_photometric_depth}}
\tablehead{
\colhead{$M_V$ limit} &
\colhead{$M_I$ limit} &
\colhead{$\sigma_{\log T_{eff}}$} &
\colhead{$\sigma_{M_{bol}}$}}
\startdata
-2.0 & -2.5 & 0.032 & 0.40 \\
-1.0 & -1.5 & 0.024 & 0.35 \\
 0.0 & -0.5 & 0.026 & 0.35 \\
 1.0 &  0.5 & 0.019 & 0.18 \\
 2.0 &  1.5 & 0.018 & 0.18 \\
 3.0 &  2.5 & 0.020 & 0.31 \\
 4.0 &  3.5 & 0.013 & 0.19 \\
\tableline
\enddata
\end{deluxetable}

\clearpage

\begin{deluxetable}{rrrr}
\tablewidth{0pt}
\tablecaption{Size of temperature and luminosity shifts as a function of filter combination.  In all three cases, the photometry reaches below the ancient main sequence turnoff.  As with Table \ref{tab_photometric_depth}, the 50\% completeness limits are given in the left two columns, while the required shifts are given in the right columns. \label{tab_filters}}
\tablehead{
\colhead{Blue limit} &
\colhead{Red limit} &
\colhead{$\sigma_{\log T_{eff}}$} &
\colhead{$\sigma_{M_{bol}}$}}
\startdata
$M_B = 5.0$ & $M_V = 4.5$ & 0.011 & 0.17 \\
$M_V = 4.0$ & $M_I = 3.5$ & 0.013 & 0.19 \\
$M_J = 3.0$ & $M_K = 2.5$ & 0.016 & 0.26 \\
\tableline
\enddata
\end{deluxetable}

\clearpage

\begin{deluxetable}{rrrr}
\tablewidth{0pt}
\tablecaption{Variation of temperature and luminosity shifts as a result of variations in photometric depth for the $B,V$ filter combination.  Columns are the same as in Table \ref{tab_photometric_depth}. \label{tab_photometric_depth_bv}}
\tablehead{
\colhead{$M_B$ limit} &
\colhead{$M_V$ limit} &
\colhead{$\sigma_{\log T_{eff}}$} &
\colhead{$\sigma_{M_{bol}}$}}
\startdata
-1.0 & -1.5 & 0.038 & 0.47 \\
 0.0 & -0.5 & 0.023 & 0.39 \\
 1.0 &  0.5 & 0.027 & 0.38 \\
 2.0 &  1.5 & 0.015 & 0.24 \\
 3.0 &  2.5 & 0.015 & 0.34 \\
 4.0 &  3.5 & 0.009 & 0.25 \\
 5.0 &  4.5 & 0.011 & 0.17 \\
\tableline
\enddata
\end{deluxetable}

\clearpage

\begin{deluxetable}{rrrr}
\tablewidth{0pt}
\tablecaption{Variation of temperature and luminosity shifts as a result of variations in photometric depth for the $J,K$ filter combination.  Columns are the same as in Table \ref{tab_photometric_depth}. \label{tab_photometric_depth_jk}}
\tablehead{
\colhead{$M_J$ limit} &
\colhead{$M_K$ limit} &
\colhead{$\sigma_{\log T_{eff}}$} &
\colhead{$\sigma_{M_{bol}}$}}
\startdata
-3.0 & -3.5 & 0.021 & 0.41 \\
-2.0 & -2.5 & 0.020 & 0.51 \\
-1.0 & -1.5 & 0.019 & 0.19 \\
 0.0 & -0.5 & 0.019 & 0.15 \\
 1.0 &  0.5 & 0.018 & 0.13 \\
 2.0 &  1.5 & 0.018 & 0.23 \\
 3.0 &  2.5 & 0.016 & 0.26 \\
\tableline
\enddata
\end{deluxetable}


\end{document}